\documentclass[twocolumn,showpacs,aps,pra]{revtex4-1}

\usepackage{amsmath,amsfonts,amssymb,bm}
\usepackage{wrapfig}
\usepackage{graphicx}
\usepackage{bbm}
\usepackage{dsfont}
\usepackage{ulem}
\usepackage{color}
\usepackage{physics}
\usepackage{natbib}
\usepackage{etoolbox}
\usepackage{lineno}
\usepackage{mathrsfs}

\begin{document}
\setlength{\columnsep}{0.8cm}
\title{Non-Markovianity of photon dynamics in a birefringent crystal}\date{\today}
\author{Kai-Hung Wang, Shih-Hsuan Chen, Yu-Cheng Lin}
\author{Che-Ming Li}
\email{cmli@mail.ncku.edu.tw}

\affiliation{Department of Engineering Science, National Cheng Kung University, Tainan 701, Taiwan}

\begin{abstract}
The way a principle system and its environment interact characterizes the non-Markovianity of the dynamics. Herein, we investigate the non-Markovian dynamics of photon polarization in a birefringent crystal. We consider the so-called ``quantity of quantum-mechanical process'' first defined by Hsieh \textit{et al.} [Sci. Rep. \textbf{7}, 13588 (2017)]. The non-Markovianity of the photon dynamics is evaluated by examining the quantity of quantum-mechanical process varying with time, and the difference between the quantity of quantum-mechanical process in a complete dynamics and that in a process composed of two subprocesses. We show that all of the processes identified as Markovian in the seminal study of Liu \textit{et al.} [Nat. Phys. \textbf{7}, 931 (2011)] can actually be identified as non-Markovian. The presented method enables us to classify non-Markovianity in the dynamical processes that are classified as Markovian by existing verification criteria. Overall, the results confirm the feasibility of performing the experimental characterization of photon dynamics using an all-optical setup and provide a useful insight into enhancing the accuracy and control of quantum system designs for quantum engineering applications.
\end{abstract}

\maketitle

\section{Introduction}

In an ideal isolated quantum system, the time evolution of the system state is governed by the Schr\"odinger equation under the postulate of quantum mechanics \cite{Nielsen&Chuang, PrincipleofQM}. In real-world scenarios, however, the physical system is inevitably affected by its environment. The dynamic behavior of such an open quantum system arises from an interaction between the system and its environment. Several important tools have been developed to describe and understand the properties of open quantum system, such as the functional integral \cite{Feynman63}, the projection operator \cite{Shibata77}, the effective modes \cite{Woods14}, the recursive
approach \cite{Gasbarri18}, and the quantum operations formalism \cite{Nielsen&Chuang, OpenQuantumSystem,Alonso}. Moreover, the system dynamics can be classified according to both the observed phenomena and dynamical map types in the mathematical formalism. Markovian and non-Markovian dynamics are recognized as important examples of distinct dynamic behavior and map structures. Accordingly, the problem of quantifying non-Markovian dynamics has attracted great attention in the literature \cite{Alonso, Colloquium, ReviewofRivas, Foundations, Hierarchy, Comparing, Markovianityandnon-MarkovianityinQandCsystems}. Compared to qualitative analysis, a quantitative description can convey more comprehensive information about the degree of non-Markovian behavior (i.e., the so-called non-Markovianity).

The first quantitative measure of non-Markovianity was proposed by Breuer, Laine, and Piilo (BLP) \cite{BLP, BLP2}, who reasoned that, if a system undergoing Markovian dynamics exhibits a monotonic decrease in the trace distance of two system states, then non-Markovianity can be defined as an increase in the trace distance. Rivas, Huelga, and Plenio (RHP) \cite{RHP} found that Markovian dynamics can cause a monotonic loss of entanglement between the system in the environment and an isolated ancilla. They thus define the non-Markovianity as an increase in the entanglement intensity, as quantified by the concurrence. From an information perspective, Markovian dynamics monotonically reduce the mutual information between the system and the ancilla. Consequently, Luo, Fu, and Song (LFS) \cite{LFS} defined the non-Markovianity as an increase in the mutual information between a system and an ancilla.

Aside from the intrinsic interest of open quantum systems, evaluating non-Markovianity is also of significant benefit in enhancing the accuracy and control of quantum system designs for quantum engineering applications. For example, the experiment of Liu \textit{et al.} \cite{Liu2011} demonstrated the feasibility for investigating the Markovian and non-Markovian dynamics of single photons in a birefringent crystal using an all-optical setup. In particular, the authors used the BLP and RHP non-Markovianity measures to classify the photon dynamics, and then experimentally demonstrated a method for controlling the dynamics transition between Markovian and non-Markovian according to these non-Markovianity measures.

Herein, we propose a method for examining  the photon dynamics in a birefringent crystal by utilizing the quantity of quantum-mechanical process. In accordance with the definition of classical process introduced by Hsieh, Chen, and Li (HCL) \cite{Our}, the  dynamics of a system are identified as classical if both the initial system states and their subsequent time evolutions can be described by classical physics \cite{Our}. Therefore, a truly quantum-mechanical process can go beyond this description. As will be shown in the following, the interactions between single photons and birefringent crystal induce the dynamical changes in the quantity of quantum-mechanical processes on photons. It will additionally be shown that such a characteristic of the system and its environment can be utilized to identify the non-Markovian dynamics for processes that cannot be identified by using existing non-Markovianity measures such as the BLP, RHP, and LFS criteria. \textit{All} of the experimental demonstrations of photon dynamics reported in Ref.~\cite{Liu2011} are strictly evaluated as non-Markovian. Finally, the conditions of the systems and experiments required to achieve a true transition from non-Markovian to Markovian dynamics are presented and discussed.

The study commences by formulating a description of the photon dynamics in terms of the process matrix. By systematically exploiting the experimental data reported in Ref.~\cite{Liu2011} to construct a faithful process matrix, we objectively analyze the non-Markovianity of the experimental processes using the HCL non-Markovian criteria. Finally, the photon dynamics classification results determined in the present study are compared with the experimental observations obtained using the BLP, RHP, and LFS non-Markovian criteria.

\begin{figure}[t]
\includegraphics[width=8.6cm]{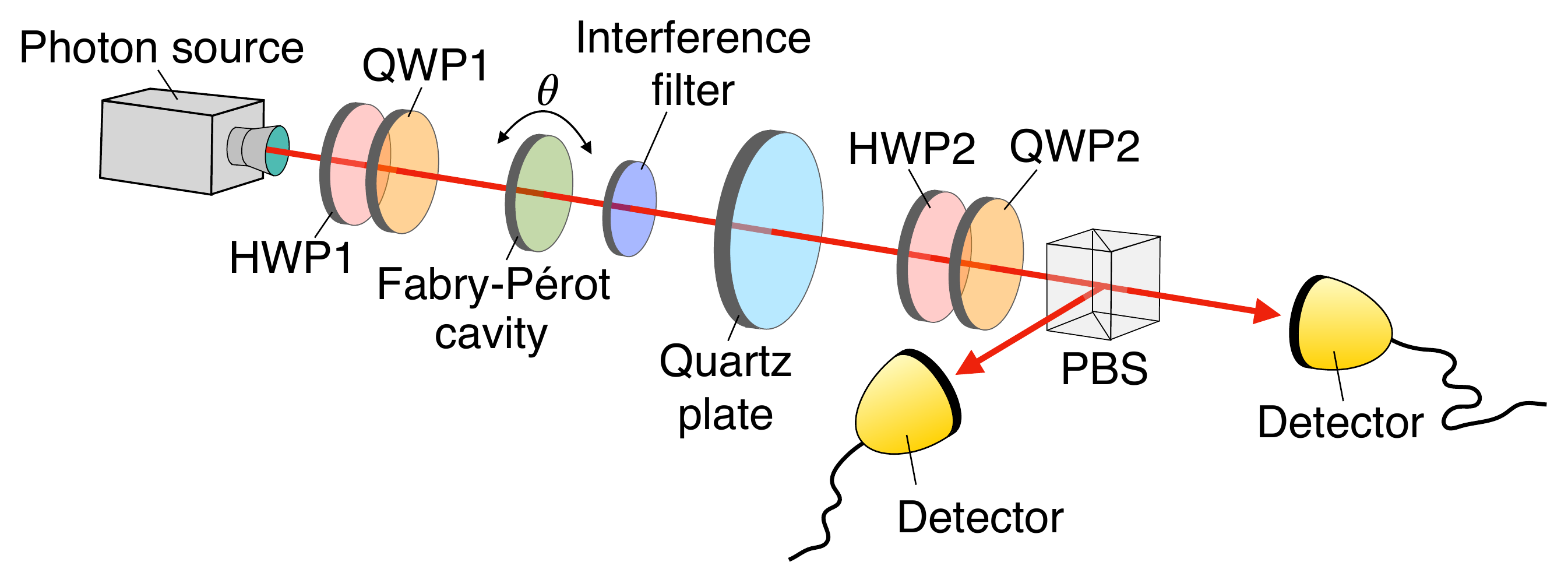}
\caption{Schematic illustration of experimental set-up used to measure the non-Markovianity of the photon dynamics in a birefringent crystal. Single photons are created from a source. The polarization degrees of freedom of the photon are initialized to an arbitrary state $\ket{\lambda}$ using a half-wave plate (HWP1) and a quarter-wave plate (QWP1). A Fabry-P\'erot (FP) cavity is then used to prepare different frequency states $\ket{\Omega^{\theta}}$ under different tilt angles $\theta$. An interference filter is placed after the FP cavity to filter out at most two transmission peaks. These two degrees of freedom of the photons are subsequently coupled together in a quartz plate, where the polarization states become dependent on the frequency states in accordance with Eq.~(\ref{unitarybirefringent}). Finally, the tomography of the polarization states is measured by a polarization analyzer composed of HWP2, QWP2, a polarizing beam splitter (PBS), and two detectors. Such a setup enables the process tomography of the photon dynamics $\chi_{t}^{\theta}$ to be measured in order to identify their non-Markovianity using the criteria given in Eqs. (\ref{tool1}) and (\ref{tool2}).}
\label{architecture}
\end{figure}

\section{Photon dynamics in birefringent crystal}

\subsection{Photonic interaction mediated by birefringent crystal}

When a photon enters a birefringent crystal (e.g., a quartz plate), the two degrees of freedom, polarization state $\ket{\lambda}$ and frequency state $\ket{\Omega^{\theta}}\equiv\int d\omega f^{\theta}(\omega)\ket{\omega}$, of the photon are coupled by the quartz plate (see Fig.~\ref{architecture}). The time evolution of the photon states in the quartz plate can be described by the following unitary transformation \cite{QP, QP2, QP3}:
\begin{equation}\label{unitarybirefringent}
U(t)\ket{\lambda}\otimes\ket{\Omega^{\theta}}=\int d\omega f^{\theta}(\omega) e^{i\omega n_{\lambda}t} \ket{\lambda}\otimes\ket{\omega},
\end{equation}
where $\lambda=H, V$ denotes the horizontal and vertical polarizations, respectively, and $n_{\lambda}$ represents the refraction index of polarization $\lambda$ of the quartz plate. In addition, the function $f^{\theta}(\omega)$ describes the amplitude in a mode with the frequency $\omega$ and can be modulated by adjusting the tilt angle $\theta$ of a Fabry-P\'erot (FP) cavity placed before the plate. The frequency distribution $|f^{\theta}(\omega)|^{2}$ satisfies the normalization condition $\int d\omega |f^{\theta}(\omega)|^{2}=1$. Since $f^{\theta}(\omega)$ is modulable, the frequency state $\ket{\Omega^{\theta}}$ can be optionally prepared by adjusting the tilt angle $\theta$ of the FP cavity as required. [Note that such a state design and creation have been experimentally demonstrated in Ref.~\cite{Liu2011} as a means of investigating the non-Markovianity of the photon dynamics shown in Eq.~(\ref{unitarybirefringent}). See Sec. \ref{Results} for a more detailed discussion.]

When we focus on the dynamical changes of the polarization states with time, the polarization degree of freedom can be considered as the principal system and the frequency degree of freedom as the environment. Equation~(\ref{unitarybirefringent}) thus describes the interaction induced between the system and its environment by the quartz plate. The interaction period of such an open system is  determined by the evolution time $t$ in Eq.~(\ref{unitarybirefringent}) and is experimentally controllable through the thickness of the quartz plate.

\subsection{Process matrix}

In order to concretely describe the photon dynamics of the principal system alone, we start by analyzing the dynamics specified by Eq. (\ref{unitarybirefringent}) using the quantum operations formalism. Let us assume that the system and environment are initially prepared in  states $\rho^{S}_{0}=\ket{\lambda}\bra{\lambda}$ and $\rho^{E}_{0}=\ket{\Omega^{\theta}}\bra{\Omega^{\theta}}$, respectively. After the interaction process in the quartz plate, $U(t)$, the final state of the system $\rho^{S}_{t}$ can be obtained by the following partial trace over the environment:
\begin{equation}\label{operation}
\begin{split}
\rho^{S}_{t}&\equiv tr_{E}[U(t) (\rho^{S}_{0}\otimes\rho^{E}_{0}) U(t)^{\dagger}] \\
&=\int_{\omega} d\omega |f^{\theta}(\omega)|^{2} \left[
    \begin{array}{cc}
    1 & 0  \\
    0 & e^{i\omega \Delta nt}  \\
    \end{array}
    \right]
        \rho^{S}_{0}
   \left [
   \begin{array}{cc}
    1 & 0  \\
    0 & -e^{i\omega \Delta nt}  \\
    \end{array}
    \right],
\end{split}
\end{equation}
where $\Delta n=n_{V}-n_{H}$ depends on the specification of the quartz plate. In other words, once the frequency distributions $|f^{\theta}(\omega)|^2$ and the change of refractive index $\Delta n$ are known, the final states $\rho^{S}_{t}$ can be determined in principle.

According to the quantum operations formalism \cite{Nielsen&Chuang, OpenQuantumSystem,Alonso}, the initial states $\rho^{S}_{0}$ and final states $\rho^{S}_{t}$ shown in Eq.~(\ref{operation}) can be associated via the following dynamical map:
\begin{equation}\label{operation2}
\chi_{t}^{\theta}: \rho^{S}_{0}\mapsto\rho^{S}_{t}.
\end{equation}
That is, the final state can be explicitly represented as
\begin{eqnarray}
\rho^{S}_{t}&\equiv&\chi_{t}^{\theta}(\rho^{S}_{0})\nonumber \\
&=&\sum_{m=1}^{4}\sum_{n=1}^{4} \chi^{\theta}_{t,mn} M_{m}\rho^{S}_{0}{M_{n}}^{\dag},\label{operatorsum}
\end{eqnarray}
where $M_{1}=I, M_{2}=X, M_{3}=-iY$ and $M_{4}=Z$. The coefficients $\chi^{\theta}_{t,mn}$ can be obtained by comparing Eqs.~(\ref{operation}) and (\ref{operatorsum}), and constitute a so-called process matrix for describing the dynamical map of the polarization states of the photons, i.e.,
\begin{equation}\label{chi1}
\begin{split}
 \chi_{t}^{\theta}
 &=
\frac{1}{4} \left[
    \begin{array}{cccc}
    2+\kappa^{\theta}(t)+\kappa^{\theta*}(t) & 0 & 0 & \kappa^{\theta}(t)-\kappa^{\theta*}(t) \\
    0 & 0 & 0 & 0 \\
    0 & 0 & 0 & 0 \\
    \kappa^{\theta*}(t)-\kappa^{\theta}(t) & 0 & 0 & 2-\kappa^{\theta}(t)-\kappa^{\theta*}(t) \\
    \end{array}
    \right].
\end{split}
\end{equation}
In Eq. (\ref{chi1}), $\kappa^{\theta}(t)$ denotes the decoherence factor and is given by the corresponding frequency distribution at tilt angle $\theta$ of the FP cavity, i.e.,
\begin{equation}\label{kappat}
\kappa^{\theta}(t)=\int{d\omega}|f^{\theta}(\omega)|^2 e^{i\omega\Delta nt}.
\end{equation}
For convenience, we use the process matrix given in Eq.~(\ref{chi1}) to refer to the physical process throughout the remainder of the text.

In practical experiments, the process matrix can be measured using a process tomography technique \cite{Nielsen&Chuang} implemented using the optical set-up shown in Fig.~\ref{architecture}. The main tomographic procedures are summarized as follows. First, the initial states of the principal system are prepared using half-wave plate 1 (HWP1) and quarter-wave plate 1 (QWP1) in order to produce four different polarization states: $\rho^{S}_{0}\in\{\ket{\lambda}\bra{ \lambda}|\lambda=H, V, +, R\}$, where $\ket{+}=(\ket{H}+\ket{V})/\sqrt{2}$ and $\ket{R}=(\ket{H}+i\ket{V})\sqrt{2}$. Following the system-environment interaction mediated by the quartz plate, the tomographic information of the final states $\rho^{S}_{t}$ is obtained using a polarization analyzer composed of HWP2, QWP2, and a polarization beam splitter (PBS). Given the measurement results for the four output states and the algorithm of process matrix, the corresponding experimental process matrices then can be systematically obtained. Ideally, these measured dynamical maps should be consistent with the theoretical maps shown in Eq.~(\ref{chi1}).

\begin{figure}[t]
\includegraphics[width=8.6cm]{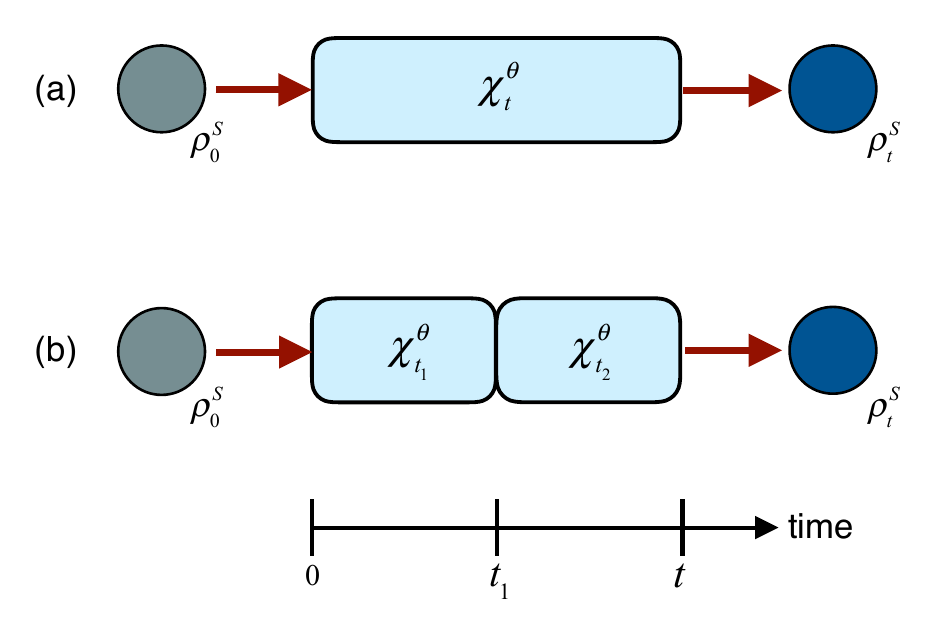}
\caption{Schematic illustration showing concept of Markovian dynamics. The initial states of the principal system $\rho^{S}_{0}$ individually undergo (a) a dynamical process $\chi^{\theta}_{t}$ and (b) a process composed of two CPTP sub-processes, $\chi^{\theta}_{t_{1}}$ and $\chi^{\theta}_{t_{2}}$ with arbitrary time $t_{1}$. If the final states $\rho^{S}_{t}$ obtained from (a) and (b) are identical for arbitrary initial states $\rho^{S}_{0}$, then $\chi^{\theta}_{t}=\chi^{\theta}_{t_2}\chi^{\theta}_{t_1}$ and the process $\chi^{\theta}_{t}$ is said to be a Markovian dynamic.}
\label{ideaofour}
\end{figure}

\section{Identifying non-Markovianity by quantifying quantum-mechanical processes}

\subsection{Markovian dynamics}
\label{MarDynamic}
As shown in Eq.~(\ref{operation2}), the evolution of the principal system $\rho^{S}_{0}$ over a time duration $t$ can be described by the process $\chi^{\theta}_{t}$. It is worth noting that, according to the quantum process formalism, $\chi^{\theta}_{t}$ is completely positive and trace preserving (CPTP) by definition, as illustrated in Eq.~(\ref{chi1}).

When the photon dynamics of the system are Markovian, the process matrix $\chi^{\theta}_{t}$ satisfies the semigroup property \cite{semigroup, DividingQuantumChannels, CPTP1, CPTP2}. In other words, $\chi^{\theta}_{t}$ can be decomposed into two separate CPTP subprocesses, $\chi^{\theta}_{t_1}$ and $\chi^{\theta}_{t_2}$, i.e.,
\begin{equation}\label{semigroup}
\chi^{\theta}_{t}=\chi^{\theta}_{t_2}\chi^{\theta}_{t_1},
\end{equation}
for arbitrary time $t_1$, where $0<t_1<t$ and $t=t_{1}+t_{2}$. (see Fig.~\ref{ideaofour}). Thus, if a dynamical map does not satisfy this Markovian condition, i.e., $\chi^{\theta}_{t}\neq\chi^{\theta}_{t_2}\chi^{\theta}_{t_1}$, $\chi^{\theta}_{t}$ can be characterized as non-Markovian.

\subsection{Classical and quantum-mechanical processes}

In order to examine the non-Markovianity of an unknown dynamics from a process perspective, let us first review the fundamental concepts of classical and quantum-mechanical processes \cite{Our}.

For a classical process, denoted as $\chi_{C}$, the initial system states can be considered as a physical object with properties which satisfy the assumption of realism and evolve according to classical stochastic theory \cite{Our}. Here, the assumption of realism indicates that the system is in a state described by a set of measurement outcomes. Moreover. the dynamics of these classical states are fully described by the transition probabilities from a specific state to a final state. If the system dynamics cannot be described at all by any classical processes, then the process is said to be a genuine quantum process (denoted by $\chi_{Q}$).

From this point of view, the system dynamics $\chi^{\theta}_{t}$ can be characterized using $\chi_{C}$ and $\chi_{Q}$. Two different methods have been proposed for giving a quantitative description of a quantum-mechanical process, namely the composition $\alpha^{\theta}_{\chi_t}$ and the robustness $\beta^{\theta}_{\chi_t}$. The composition $\alpha^{\theta}_{\chi_t}$ describes the minimum quantity of quantum processes that can be found in $\chi^{\theta}_{t}$, i.e.,
\begin{equation}\label{tool1}
\chi^{\theta}_{t}=\alpha^{\theta}_{\chi_t}\chi_{Q}+(1-\alpha^{\theta}_{\chi_t})\chi_{C}.
\end{equation}
By contrast, the robustness $\beta^{\theta}_{\chi_t}$ shows the ability of $\chi^{\theta}_{t}$ to tolerate the minimum amount of noise to become a classical process, i.e.,
\begin{equation}
\frac{\chi^{\theta}_{t}+\beta^{\theta}_{\chi_t}\chi'}{1+\beta^{\theta}_{\chi_t}}=\chi_{C},
\end{equation}
where $\chi'$ denotes the process noise \cite{Our}. For example, if $\chi^{\theta}_{t}$ is unitary, we have $\alpha^{\theta}_{\chi_t}=1$ and $\beta^{\theta}_{\chi_t}\simeq0.4641$, which implies that it cannot be described by any classical process. The $\alpha^{\theta}_{\chi_t}$ and $\beta^{\theta}_{\chi_t}$ for process can be considered as an extension of the composition \cite{composition} and robustness \cite{robustness} for describing state characteristics.

\subsection{The HCL non-Markovian criteria}

As shown in Ref. \cite{Our}, $\alpha$ and $\beta$ should monotonically decrease with time for a Markovian process. Hence, if an increasing result from $\chi^{\theta}_{t}$ is observed, then the process can be verified as non-Markovian. In other words, the non-Markovianity of $\chi^{\theta}_{t}$ can be evaluated by integrating the positive derivative of $\alpha^{\theta}_{t}$ or $\beta^{\theta}_{\chi_t}$ with respect to time, i.e.,
\begin{equation}\label{tool1}
\begin{split}
\mathscr{W}^{\theta}_{\alpha^{\theta}_{\chi_t}}\equiv\int_{0 ;\frac{d\alpha}{dt}>0}^{t}\dot{\alpha^{\theta}_{\chi_t}}dt >0,\\
\mathscr{W}^{\theta}_{\beta^{\theta}_{\chi_t}}\equiv\int_{0 ;\frac{d\beta}{dt}>0}^{t}\dot{\beta^{\theta}_{\chi_t}}dt >0,
\end{split}
\end{equation}
where the superscripts of $\mathscr{W}^{\theta}_{\alpha^{\theta}_{\chi_t}}$ and $\mathscr{W}^{\theta}_{\beta^{\theta}_{\chi_t}}$ indicate that the angle of the FP cavity $\theta$ is chosen in the experiment.

If, on the other hand, $\chi^{\theta}_{t}$ is Markovian, there exists no difference between $\alpha^{\theta}_{\chi_{t}}$ and $\alpha^{\theta}_{\chi_{t_{2}} \chi_{t_{1}}}$ and $\beta^{\theta}_{\chi_{t}}$ and $\beta^{\theta}_{\chi_{t_{2}} \chi_{t_{1}}}$ according to the Markovian condition given in Eq.~(\ref{semigroup}) (and shown also in Fig.~\ref{ideaofour}). Here the subscripts of $\alpha^{\theta}_{\chi_{t}}$ and $\alpha^{\theta}_{\chi_{t_{2}} \chi_{t_{1}}}$ ($\beta^{\theta}_{\chi_{t}}$ and $\beta^{\theta}_{\chi_{t_{2}} \chi_{t_{1}}}$) indicate that the composition $\alpha$ (robustness $\beta$) is derived from a single process $\chi_{t}$ and composite process $\chi_{t_{2}} \chi_{t_{1}}$, respectively. An invalidation of this consistency reveals that the process is non-Markovian. In other words, non-Markovian dynamics are identified when
\begin{equation}\label{tool2}
\begin{split}
\mathscr{N^{\theta}_{\alpha}}\equiv|\alpha^{\theta}_{\chi_{t}}-\alpha^{\theta}_{\chi_{t_{2}} \chi_{t_{1}}}|>0,\\
\mathscr{N^{\theta}_{\beta}}\equiv|\beta^{\theta}_{\chi_{t}}-\beta^{\theta}_{\chi_{t_{2}} \chi_{t_{1}}}|>0.
\end{split}
\end{equation}

\section{Non-Markovian dynamics of photon polarization state}
\label{Results}

\subsection{Experimental results of Liu \textit{et al.} \cite{Liu2011}}
\label{Previous experiment}

In the experimental work of Liu \textit{et al.}, the non-Markovianity of the photon dynamics is first classified using the BLP criterion based on the dynamical changes of the trace distance between two system states. Consider two system initial states, $\rho_{0,1}^{S}$ and $\rho_{0,2}^{S}$, with a trace distance $D_{0}^{S}=\|\rho_{0,1}^{S}-\rho_{0,2}^{S}\|/2$ between them, and two corresponding final states, $\rho_{t,1}^{S}$ and $\rho_{t,2}^{S}$, with $D_{t}^{S}={ \|\rho_{t,1}^{S}-\rho_{t,2}^{S}\|}/2$. If $D^{S}_{t}>D^{S}_{0}$, the dynamics of $\chi^{\theta}_{t}$ are classified as non-Markovian according to the BLP criterion. (See Appendix~\ref{BLPc} for a more detailed review.)

\begin{figure}[t]
\includegraphics[width=8.6cm]{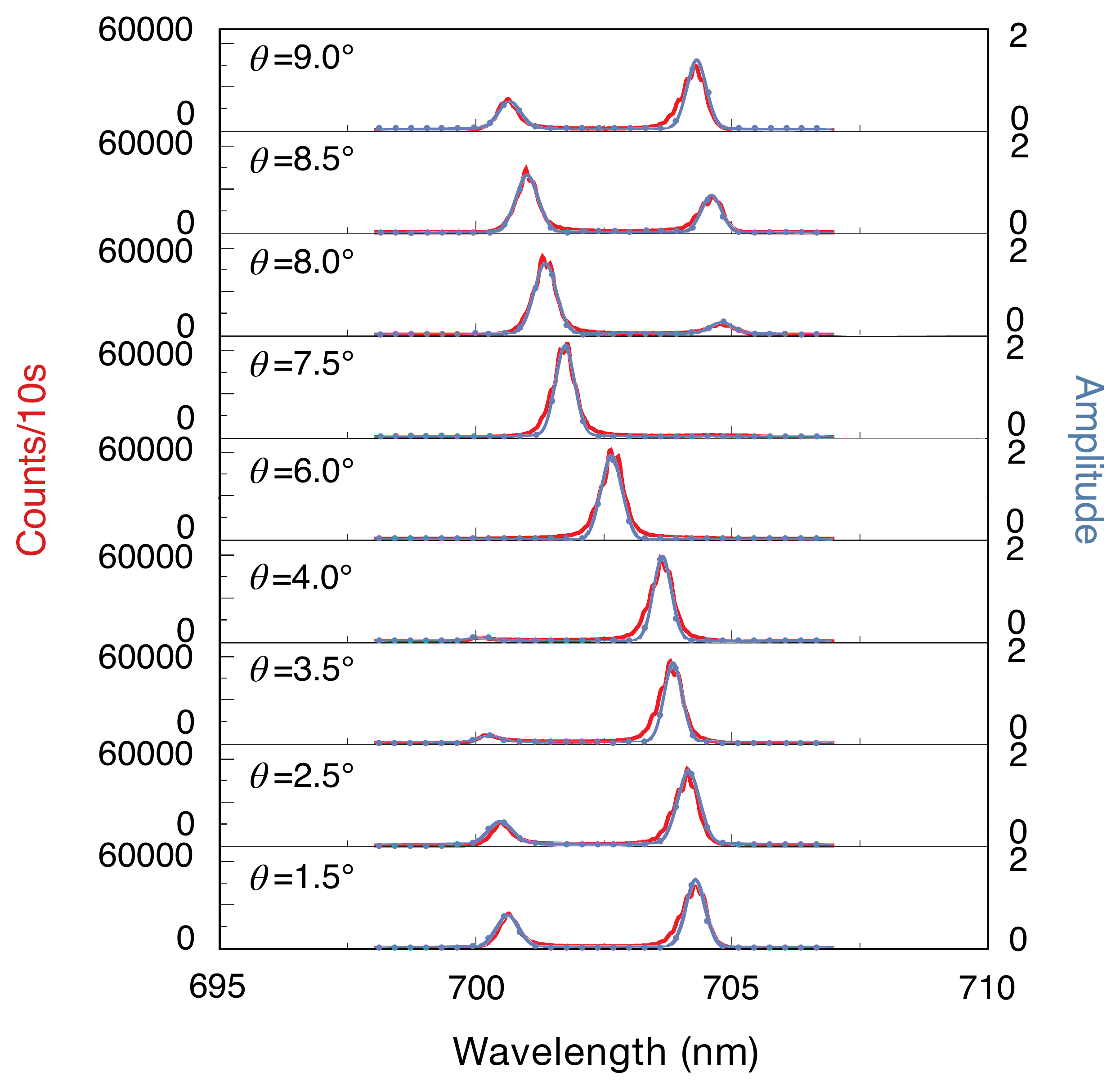}
\caption{Simulated frequency spectrums of the initial state for various values of the tilt angle $\theta$. Note that the simulated spectrums (blue-dotted lines) are constructed using the experimental data measured by a monochromator (red lines) reported in Ref.~\cite{Liu2011}. Note also that the count rate of the monochromator corresponds to the amplitude which is an introduced parameter used in the frequency distribution function.}
\label{ReconstructSpectrum}
\end{figure}

\begin{figure}[t]
\includegraphics[width=8.6cm]{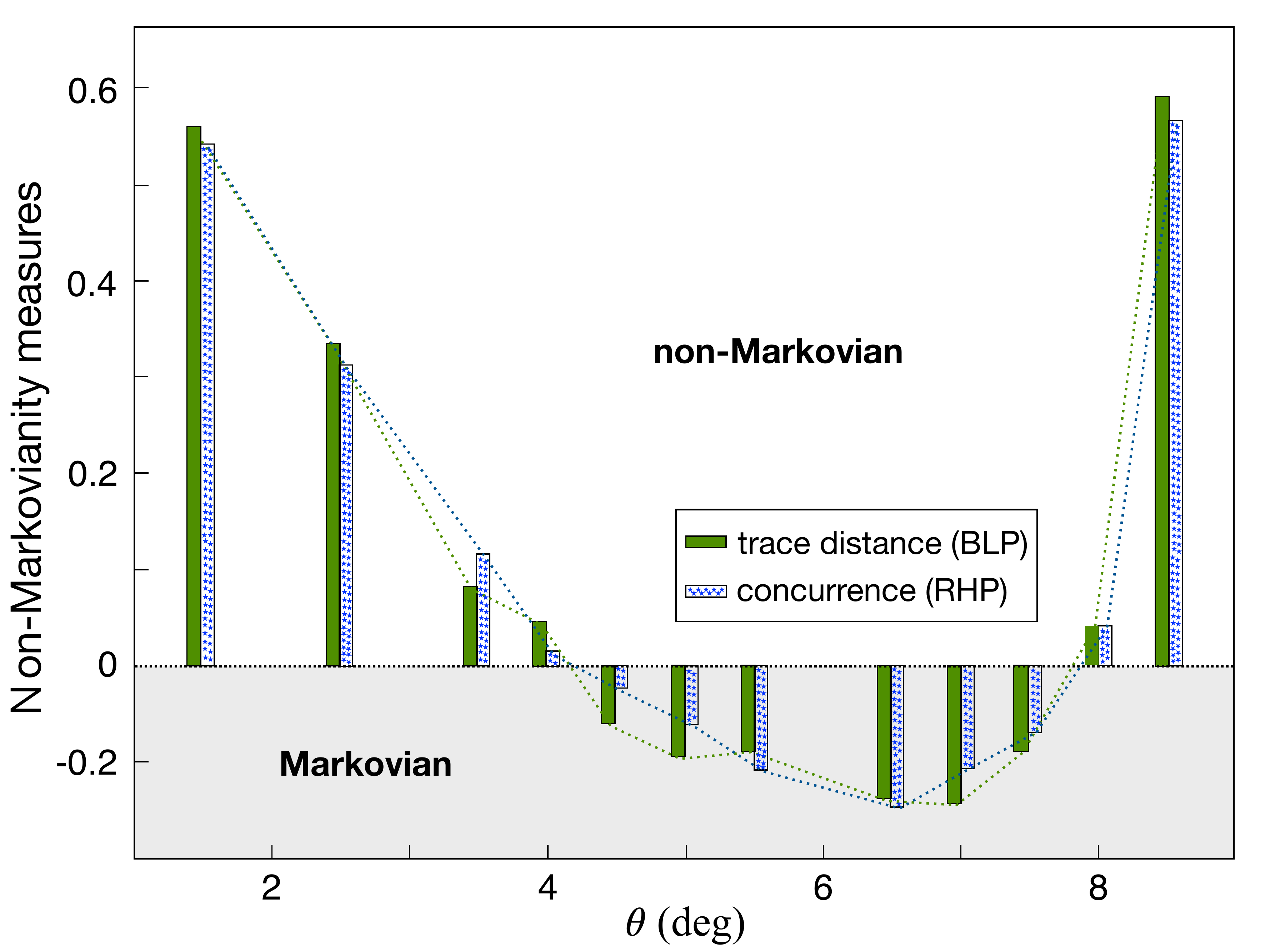}
\caption{Non-Markovianity of the photon dynamics reported in Ref.~\cite{Liu2011}. The green and blue bars show the non-Markovianity identified by the BLP criterion and RHP criterion, respectively. Note that, when $4.0^{\circ}<\theta<8.0^{\circ}$, the photon dynamics are Markovian, as characterized by both the BLP and RHP criteria.}
\label{changeofTDandCon}
\end{figure}

In the experiments reported in Ref.~\cite{Liu2011}, the two system states are initially prepared as $\rho^{S}_{0,1}=(\ket{H}+\ket{V})/\sqrt{2}$ and $\rho^{S}_{0,2}=(\ket{H}-\ket{V})/\sqrt{2}$ using a HWP. Note that the experimental setup is similar to that shown in Fig.~\ref{architecture}. In particular, specific tilt angles $\theta$ of the FP cavity are set to create different initial states of the environment. A 4 nm FWHM (full width at half maximum) interference filter is used to filter out at most two transmission peaks. The frequency distribution corresponding to each value of $\theta$ is then measured using a monochromator, as illustrated in Fig.~\ref{ReconstructSpectrum}. After interacting with the environment in the quartz plate as described in Eq.~(\ref{unitarybirefringent}), the final states of the principal system, i.e., $\rho^{S}_{t, 1}$ and $\rho^{S}_{t, 2}$, respectively, are measured tomographically using a polarization analyzer. The classification of the photon dynamics reported in Ref.~\cite{Liu2011} is shown in Fig.~\ref{changeofTDandCon}. As shown, for $4.0^{\circ}<\theta<8.0^{\circ}$, the photon dynamics are identified as Markovian using the BLP criterion.

The experimental work of Liu \textit{et al.} \cite{Liu2011} also demonstrates the RHP non-Markovian criterion, and compares the observation results with those obtained using the BLP criterion. The related experiment prepares a maximally entangled state of the principal system and an ancilla which is isolated from the environment, as $\rho^{SA}_{0}=\ket{\psi^{SA}_{0}}\bra{\psi^{SA}_{0}}$, in which $\ket{\psi_{0}^{SA}}=(\ket{H^SH^A}+\ket{V^SV^A})/\sqrt{2}$, and the superscript $A$ denotes the ancilla. The concurrence of this initial state is $C(\rho^{SA}_{0})=1$ \cite{concurrence}. Since the principal system is subject to the dynamical process $\chi_{t}^{S}$, the state of the composite system is given by $\rho^{SA}_{t}=\chi_{t}^{S}\otimes\mathbb{I}_{t}^{A}(\rho^{SA}_{0}$), where $\mathbb{I}_{t}^{A}$ denotes an identity process operating on the ancilla. In the experiment, the final state $\rho^{SA}_{t}$ can be measured using two-photon state tomography, and hence the $C(\rho^{SA}_{t})$ can directly be determined. A Markovian process causes $C(\rho^{SA}_{t})$ to monotonically decrease with time $t$. The experimental results classified by the RHP criterion are shown in Fig.~\ref{changeofTDandCon}, and indicate that the photon dynamics for $4.0^{\circ}<\theta<8.0^{\circ}$ are Markovian. In other words, the identification results are identical to those obtained using the BLP criterion. (A more detailed discussion regarding concurrence and the RHP criterion is presented in Appendix~\ref{RHPc}.)

\subsection{Non-Markovianity identified by the criteria (\ref{tool1}) and (\ref{tool2})}
\label{ourresult}

\subsubsection{Construction of the experimental process matrix $\chi^{\theta}_{t}$}
\label{Constructionoftheexperimentalprocessmatrix}

In order to identify the non-Markovianity of $\chi^{\theta}_{t}$ in Eq.~(\ref{chi1}) using the criteria (\ref{tool1}) and (\ref{tool2}), it is necessary to completely know the frequency distribution $|f^{\theta}(\omega)|^2$ and refraction index difference $\Delta n$ in Eq.~(\ref{kappat}). Here, in order to objectively compare the present results with those of Liu \textit{et al.} \cite{Liu2011} (as will be shown in Sec. \ref{comparisonsanddiscussions}), $|f^{\theta}(\omega)|^2$ and $\Delta n$ are derived directly from their experimental report data.

Notably, there are no exact functions available for describing the frequency distributions in Ref.~\cite{Liu2011}. Thus, it is necessary to construct the frequency distributions from scratch by using a frequency model to fit the measured frequency spectrums reported in the related experiment. In practice, the frequency distributions can be modeled as the sum of two Gaussian distributions with frequency centers of $\omega_{c1}$ and $\omega_{c2}$, respectively, and standard deviations of $\sigma_{1}$ and $\sigma_{2}$, i.e.,
\begin{equation}\label{freq}
|f^{\theta}(\omega)|^2=\sum_{j=1} ^2 \frac{a_j}{a_1+a_2}  \frac{1}{\sigma_{j}\sqrt{2\pi}} \exp({- \frac{(\omega-\omega_{cj})^2}{2\sigma_{j}^2}}),
\end{equation}
where $a_1/(a_1+a_2)$ and $a_2/(a_1+a_2)$ are the amplitudes corresponding to the sub-distributions with centers $\omega_{c1}$ and $\omega_{c2}$, respectively. Function $|f^{\theta}(\omega)|^2$ in Eq.~(\ref{kappat}) for each given angle $\theta$ can then be determined by fitting the measured frequency spectrums with Eq.~(\ref{freq}). As illustrated in Fig.~\ref{ReconstructSpectrum}, the simulated frequency distributions are highly consistent with the experimental spectrums given the fitting parameters values $\sigma_{1}$, $\sigma_{2}$, $a1$, $a2$, $\omega_{c1}$ and $\omega_{c2}$ listed in Table.~\ref{FreAnalysis}.
\begin{table}[t]
\caption{Parameters used in fitting the experimental frequency spectrums of Liu \textit{et al.} \cite{Liu2011}. The frequency distributions $|f^{\theta}(\omega)|^2$ created using Eq.~(\ref{freq}) are consistent with the experimental spectrums reported in Ref. \cite{Liu2011}. Given the frequency distributions generated from Eq.~(\ref{freq}) and $\Delta n$, Eqs.~(\ref{chi1}) and (\ref{chi2}) can be used to specify the photon dynamics for each considered $\theta$.\label{FreAnalysis}}
\begin{ruledtabular}
\begin{tabular}{ccccccc}
$\theta(^{\circ})$ & $\omega_{c1}\textrm{(nm)}$ & $\omega_{c2}\textrm{(nm)}$ & $a_{1}$ & $a_{2}$ & $\sigma_{1}\textrm{(nm)}$ & $\sigma_{2}\textrm{(nm)}$ \\
\hline
1.5 & 700.608 & 704.286 & 0.787 & 1.455 & 0.185 & 0.212 \\
2.5 & 700.476 & 704.153 & 0.545 & 1.636 & 0.212 & 0.225 \\
3.5 & 700.238 & 703.836 & 0.182 & 1.787 & 0.172 & 0.225 \\
4.0 & 700.079 & 703.651 & 0.901 & 1.848 & 0.172 & 0.212 \\
6.0 & 702.672 & {---} & 1.848 & {---} & 0.198 & {---} \\
7.5 & 701.720 & {---} & 1.909 & {---} & 0.185 & {---} \\
8.0 & 701.349 & 704.788 & 1.636 & 0.273 & 0.212 & 0.212 \\
8.5 & 701.005 & 704.603 & 1.333 & 0.758 & 0.185 & 0.212 \\
9.0 & 700.635 & 704.286 & 0.667 & 1.545 & 0.185 & 0.212 \\
\end{tabular}
\end{ruledtabular}
\end{table}

To determine the experimental value of $\Delta n$, the frequency distributions generated in Eq.~(\ref{freq}) are used to fit the oscillation periods of the change of the trace distance and the concurrence in the reported experimental results. (See Fig. 3 in Ref.~\cite{Liu2011}.) A value of $\Delta n=0.0115$ is found to achieve the best fit between the simulation results and the experimental results. Thus the photon dynamics $\chi^{\theta}_{t}$ in Eq.~(\ref{chi1}) can be derived from the experimental results and data shown in Ref.~\cite{Liu2011}.

In order to confirm the faithfulness of the process matrix $\chi^{\theta}_{t}$ derived in Eq.~(\ref{chi1}), the BLP and RHP  non-Markovianity measures are also applied to examine $\chi^{\theta}_{t}$. The simulation results are again consistent with the experimental observations in Ref.~\cite{Liu2011} (Fig.~\ref{changeofTDandCon}). (See Figs. \ref{resultofBLP} and \ref{resultofRHP} in Appendixes \ref{BLPc} and \ref{RHPc} for a more detailed discussions.) Note that, when $\theta=6.0^{\circ}$ and $7.5^{\circ}$, neither the experimental results nor the simulation results identify non-Markovianity when using the BLP and RHP criteria. The same result is found from all the derived photon dynamics obtained under the fitting parameters listed in Table~\ref{FreAnalysis}. Hence the process matrices constructed with the frequency model in Eq.~(\ref{freq}) provided a reliable means of investigating the non-Markovianity of photon dynamics.

\subsubsection{Identifying non-Markovianity using the criterion (\ref{tool1})}

In order to compare the experimental results shown in Fig.~\ref{changeofTDandCon}, we first consider the photon dynamics $\chi^{\theta}_{t}$ for $\theta=6.0^{\circ}$ and $7.5^{\circ}$, corresponding to the Markovian region in Fig.~\ref{changeofTDandCon}. The  quantities of $\alpha$ and $\beta$ of the dynamical processes $\chi^{6.0}_{t}$ and $\chi^{7.5}_{t}$ both monotonically decrease with  $t$, as shown in Fig.~\ref{resultofour}. The photon dynamics $\chi^{6.0}_{t}$ and $\chi^{7.5}_{t}$ therefore do not satisfy the criterion in Eq.~(\ref{tool1}) for the non-Markovianity.

However, let us now examine the photon dynamics corresponding to the transition region of non-Markovian and Markovian dynamics in Fig.~\ref{changeofTDandCon}. As illustrated in Fig.~\ref{resultofour2}, the values of $\alpha^{6.0}_{\chi_{t}}$ and $\beta^{6.0}_{\chi_{t}}$ do not monotonically decrease. Thus, according to the criterion given in Eq.~(\ref{tool1}), the photon dynamics $\chi^{4.0}_{t}$ and $\chi^{8.0}_{t}$ are identified as non-Markovian.

\subsubsection{Identifying non-Markovianity using the criterion (\ref{tool2})}

\begin{figure}
\includegraphics[width=8.6cm]{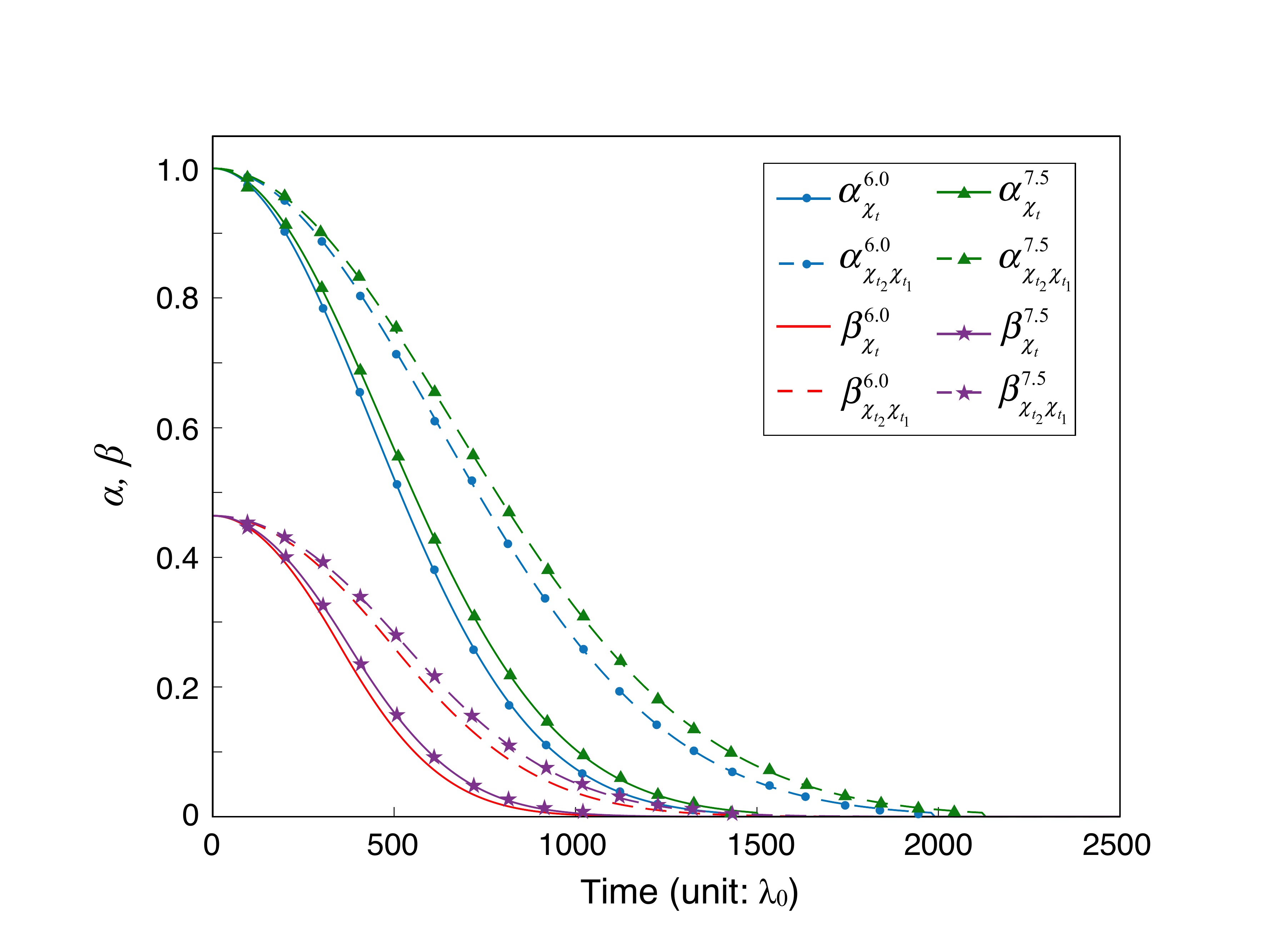}
\caption{Changes of composition and robustness over time for $\chi^{\theta}_{t}$ and $\chi_{t_{2}}^{\theta}\chi_{t_{1}}^{\theta}$, where $\theta=6.0^{\circ}$ and $7.5^{\circ}$. The $\lambda_{0}$ is 702 nm, which denotes the effective path difference corresponding to Ref.~\cite{Liu2011}. Since $\alpha^{6.0}_{\chi_{t}}$ ($\beta^{6.0}_{\chi_{t}}$) monotonically decays with time, the system dynamics cannot be identified as non-Markovian using Eq. (\ref{tool1}). However, the existence of differences between $\alpha_{\chi_{t}}^{6.0}$ and $\alpha_{\chi_{t_2}\chi_{t_1}}^{6.0}$ ($\beta_{\chi_{t}}^{6.0}$ and $\beta_{\chi_{t_2} \chi_{t_1}}^{6.0}$) indicates the non-Markovianity of the photon dynamics in accordance with the criterion in Eq.~(\ref{tool2}). These features can be seen when characterizing $\chi_{t}^{7.5}$. Thus, $\chi^{6.0}_{t}$ and $\chi^{7.5}_{t}$ are both identified as non-Markovian by the criterion (\ref{tool2}). Note, however, that they are classified as Markovian by the BLP and RHP criteria in Ref.~\cite{Liu2011}.}
\label{resultofour}
\end{figure}

\begin{figure}[t]
\includegraphics[width=8.6cm]{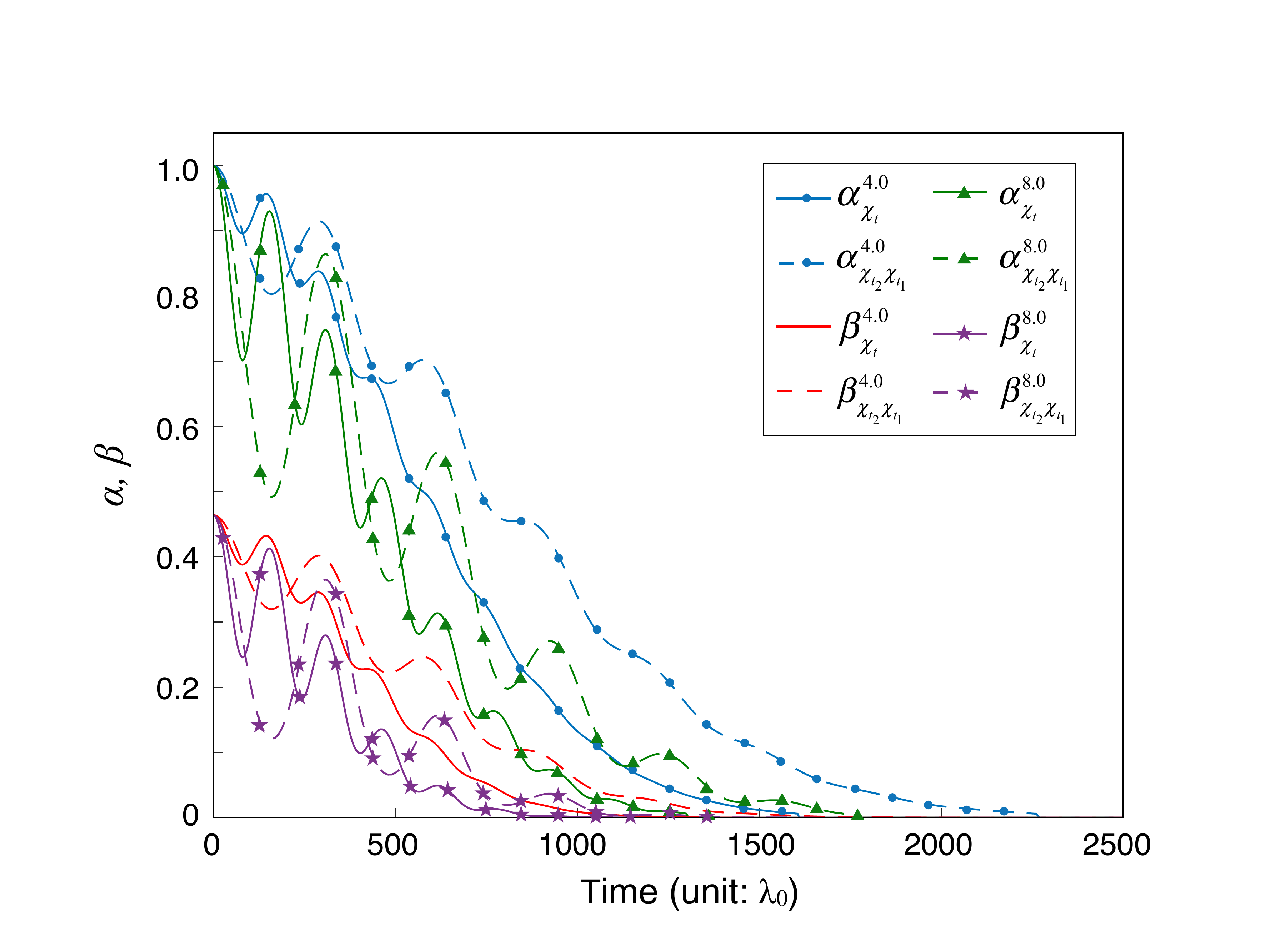}
\caption{Changes of composition and robustness over time for $\chi^{\theta}_{t}$ and $\chi_{t_{2}}^{\theta}\chi_{t_{1}}^{\theta}$, where $\theta=4.0^{\circ}$ and $8.0^{\circ}$. The results show that the photon dynamics not only satisfy the non-Markovian criterion in Eq.~(\ref{tool1}), but also the criterion in Eq.~(\ref{tool2}). Hence, $\chi_{t}^{4.0}$ and $\chi_{t}^{8.0}$ are thus both identified as non-Markovian. This result is consistent with that obtained from the BLP and RHP criteria in Ref.~\cite{Liu2011}.}
\label{resultofour2}
\end{figure}

To examine the non-Markovianity of $\chi^{\theta}_{t}$ using the non-Markovian criterion in Eq.~(\ref{tool2}), we first need to compare $\alpha$ and $\beta$ of $\chi^{\theta}_{t}$ with those of the composite process $\chi_{t_{2}}^{\theta}\chi_{t_{1}}^{\theta}$. Here, we choose $t_{1}=t_{2}={t}/2$ (see Fig.~\ref{ideaofour}). The dynamical map consisting of the two sub-processes can be formulated as:
\begin{widetext}
\begin{equation}\label{chi2}
 \chi_{t_{2}}^{\theta} \chi_{t_{1}}^{\theta}=\frac{1}{4} \left[
    \begin{array}{cccc}
    2+{\kappa^{\theta}(\frac{t}{2})}^2+{\kappa^{\theta*}(\frac{t}{2})}^2 & 0 & 0 & {\kappa^{\theta}(\frac{t}{2})}^2-{\kappa^{\theta*}(\frac{t}{2})}^2 \\
    0 & 0 & 0 & 0 \\
    0 & 0 & 0 & 0 \\
    {\kappa^{\theta*}(\frac{t}{2})}^2-{\kappa^{\theta}(\frac{t}{2})}^2 & 0 & 0 & 2-{\kappa^{\theta}(\frac{t}{2})}^2-{\kappa^{\theta*}(\frac{t}{2})}^2 \\
    \end{array}
    \right].
\end{equation}
\end{widetext}
Figures \ref{resultofour} and \ref{resultofour2} show that there exist clear differences between $\alpha$ and $\beta$ for $\chi^{\theta}_{t}$ and those for $\chi_{t_{2}}^{\theta}\chi_{t_{1}}^{\theta}$ for almost all of the time periods. The results thus provide clear evidence of non-Markovian dynamics according to Eq. (\ref{tool2}). In other words, the photon dynamics for $\theta=4.0^{\circ}, 6.0^{\circ}, 7.5^{\circ}$ and $8.0^{\circ}$ are classified as non-Markovian.

\subsection{Discussions}
\label{comparisonsanddiscussions}

\subsubsection{Comparisons}
\label{comparison}
Figure \ref{resultofour} shows that when the non-Markovianity of a dynamical process is characterized by examining the changes in the amount of quantum process over time, the non-Markovianity of the photon dynamics can be successfully identified using the criterion given in Eq.~(\ref{tool2}). By contrast, such non-Markovianity cannot be identified using the BLP or RHP criterion. Indeed, all of the photon dynamics $\chi^{\theta}_{t}$ for $4.0^{\circ}<\theta<8.0^{\circ}$ classified as Markovian by the BLP and RHP criteria are identified as non-Markovian by the criterion~(\ref{tool2}). However, for all of the other tilt angles considered in Table~\ref{FreAnalysis}, the corresponding photon dynamics are classified as non-Markovian by all the non-Markovian criteria.

According to the LFS criterion, all of the photon dynamics for the cases considered in Table.~\ref{FreAnalysis} are non-Markovian other than those for $\chi^{6.0}_{t}$ and $\chi^{7.5}_{t}$. It is noted that this result is consistent with that obtained using the BLP and RHP non-Markovian criteria. For each setting shown in Table~\ref{FreAnalysis}, the simulation results obtained using the LFS non-Markovian criterion for each of the parameter settings shown in Table.~\ref{FreAnalysis} are presented in Appendix \ref{LFSc}.

\subsubsection{Switching non-Markovian photon dynamics to Markovian photon dynamics}
According to the comparisons in Sec. \ref{comparison}, the criterion (\ref{tool2}) can be used to classify non-Markovianity of the photon dynamical processes that are classified as Markovian by the BLP, RHP, and LFS criteria. As such, it provides a useful tool for realizing more accurate system-environment quantum engineering designs. This benefit is illustrated in the following, in which the all-optical set-up in Fig.~\ref{architecture} is used to switch the photon dynamics from non-Markovian to Markovian.

Let us state by considering a single frequency distribution centered at $\omega_{s}$ rather than the two Gaussian distributions considered in Eq.~(\ref{freq}). Term $\kappa^{\theta}(t)$ in Eqs. (\ref{chi1}) and (\ref{chi2}) is found to be $e^{i\omega_{s}\Delta nt}$ following the Fourier transformation. In other words, $\chi^{\theta}_{t}$ describes a unitary transformation. This satisfies the Markovian condition in Eq.~(\ref{semigroup}). Thus it can be inferred that, as the value of $\sigma$ reduces, the photon dynamics approach a Markovian process. Accordingly, the transition of the photon dynamics from non-Markovian to Markovian, as identified by criterion in Eq.~(\ref{tool2}), can be simulated by modifying the value of $\sigma$ in Eq.~(\ref{freq}).

To compare the present observations with the previous experiment in Ref.~\cite{Liu2011}, consider the case of $\chi_{t}^{6.0}$, for example, which is the most Markovian part characterized by the BLP and RHP non-Markovian criteria (see Fig. \ref{changeofTDandCon}). Let the evolution time be chosen as $t=160\lambda_0$, where this choice depends on the evolution time of the photons in the experiment (see Fig. 3 in Ref. \cite{Liu2011}). In classifying the photon dynamics using Eq.~(\ref{tool2}), both the single process and the related subprocesses are of interest. Thus, the time intervals are divided into two equal parts (e.g., as shown in Fig.~\ref{ideaofour}); resulting in $t_{1}=80\lambda_{0}$. A minimum resolution of $1\%$ is assumed for both $\mathscr{N^{\rm{6.0}}_{\alpha}}$ and $\mathscr{N^{\rm{6.0}}_{\beta}}$ in order to account for potential errors in the experiments. In other words, if both $\mathscr{N^{\rm{6.0}}_{\alpha}}$ and $\mathscr{N^{\rm{6.0}}_{\beta}}<1\%$ during evolution time $t=160\lambda_{0}$, the changes of $\alpha$ and $\beta$ of the original process and those of the two sub-processes, respectively, are recognized as being the same.

\begin{figure}[h]
\includegraphics[width=8.6cm]{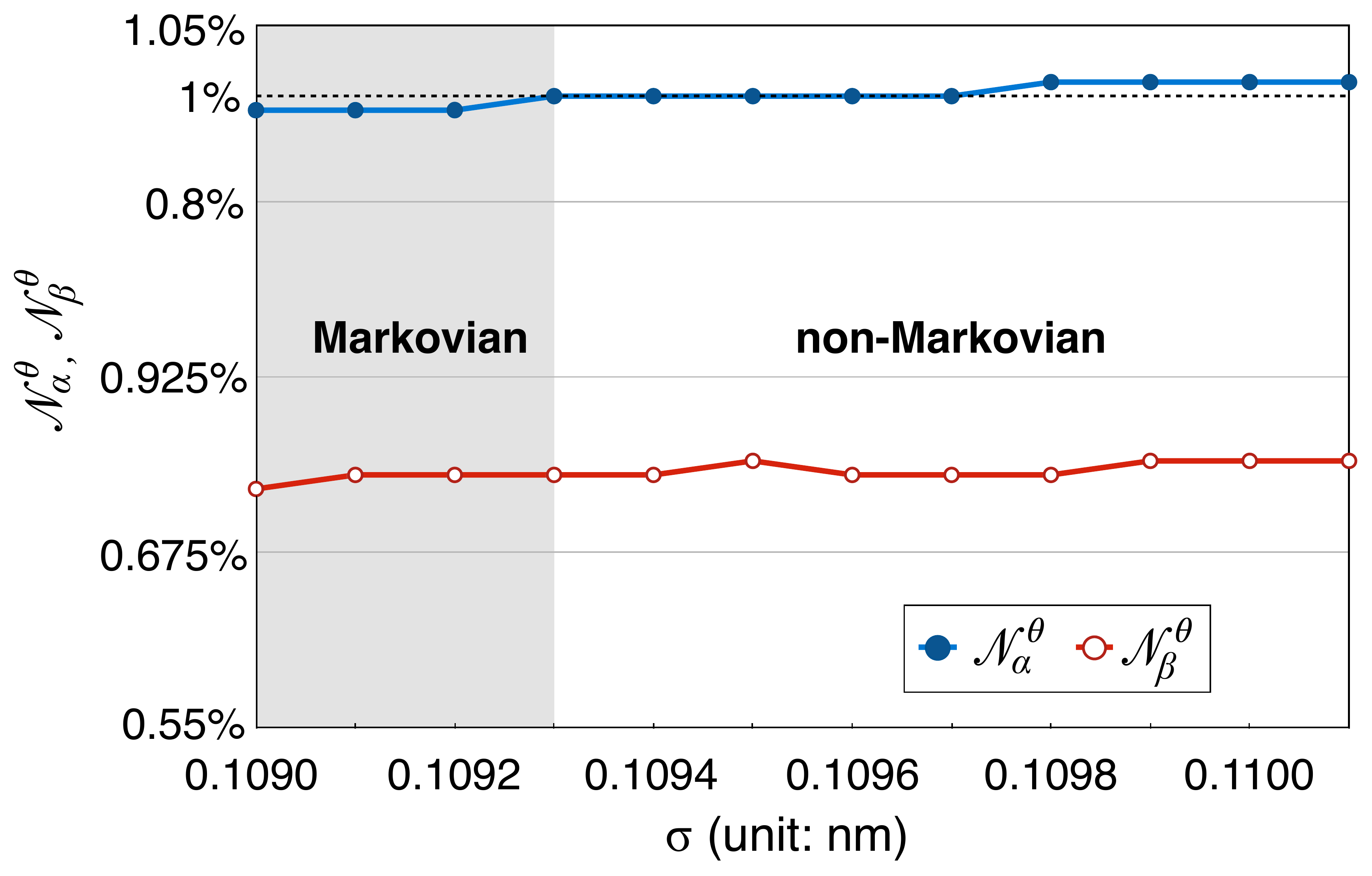}
\caption{$\mathscr{N^{\rm{6.0}}_{\alpha}}$ and $\mathscr{N^{\rm{6.0}}_{\beta}}$ with different $\sigma$ at $t=160\lambda_{0}$. By adjusting the value of $\sigma$ in Eq.~(\ref{freq}), the photon dynamics can be switched from non-Markovian (white area) to Markovian (gray area) when $\sigma<0.1093$ (FWHM $<0.2574$), as identified by Eq.~(\ref{tool2}).}
\label{trans}
\end{figure}

Figure \ref{trans} shows the values of $\mathscr{N^{\rm{6.0}}_{\alpha}}$ and $\mathscr{N^{\rm{6.0}}_{\beta}}$ at $t=160\lambda_{0}$ given different choices of $\sigma$. The results confirm that $\mathscr{N^{\rm{6.0}}_{\alpha}}$ and $\mathscr{N^{\rm{6.0}}_{\beta}}$ are both less than $1\%$ when $\sigma<0.1093$. In other words, the photon dynamics $\chi^{6.0}_{160\lambda_0}$ are changed to Markovian when $\sigma<0.1093$, as evaluated  by the criterion in Eq.~(\ref{tool2}).

\begin{figure}[t]
\includegraphics[width=8.6cm]{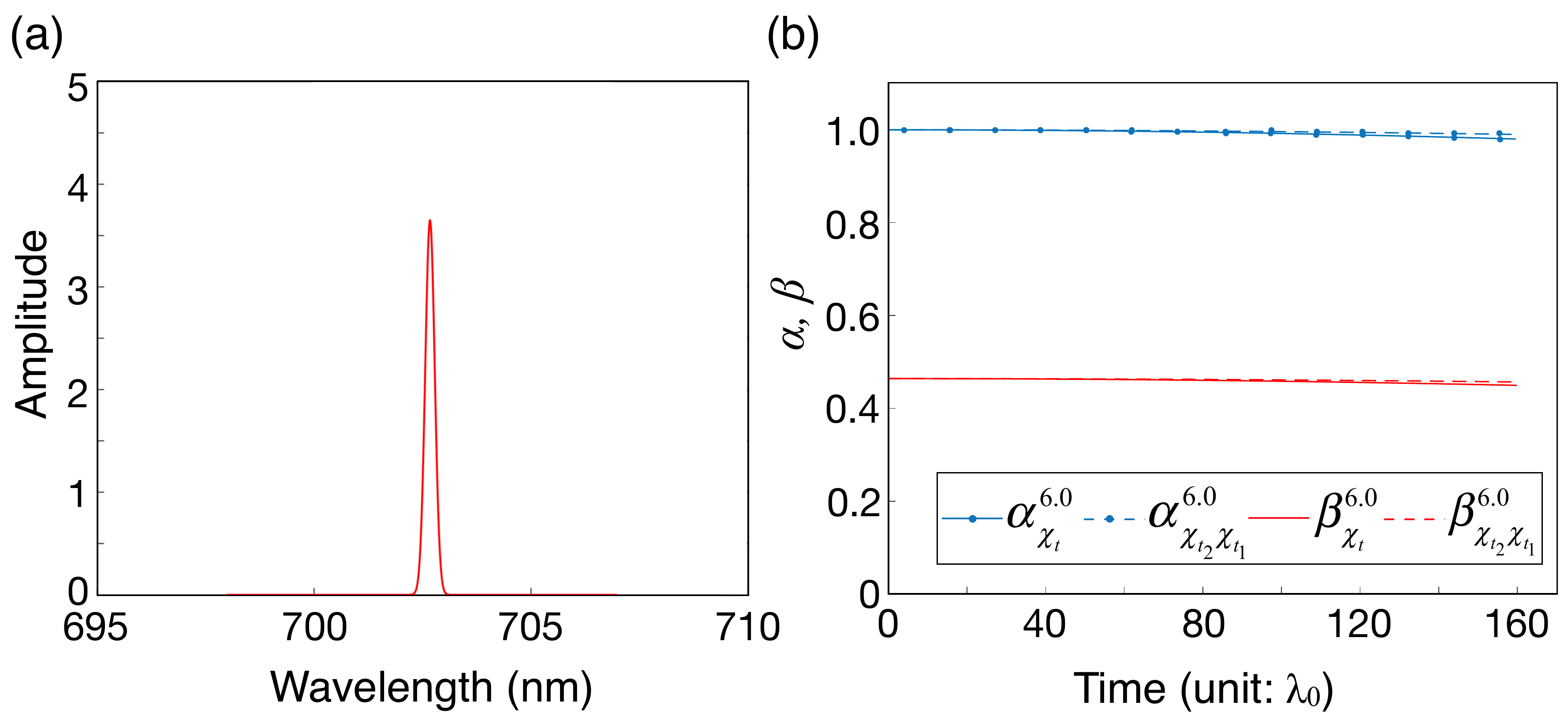}
\caption{Conditions and evidence for switching $\chi^{6.0}_{t}$ to the Markovian dynamics under the criterion (\ref{tool2}). (a) The conditions of $\omega_{c}=702.6720$ nm and $\sigma=0.1092$ nm constituting a single-peak frequency spectrum are used to switch $\chi^{6.0}_{t}$ to a Markovian process. (b) The evidence of switching $\chi^{6.0}_{t}$ to a Markovian process. The $\alpha_{\chi_{t}}^{6.0}$ and $\alpha_{\chi_{t_2}\chi_{t_1}}^{6.0}$ ($\beta_{\chi_{t}}^{6.0}$ and $\beta_{\chi_{t_2}\chi_{t_1}}^{6.0}$) monotonically decrease over time from $t=0$ to $t=160\lambda_{0}$. There exists a difference which is less than 1$\%$ between them ($\mathscr{N^{\rm{6.0}}_{\alpha}}=0.0099$ and $\mathscr{N^{\rm{6.0}}_{\beta}}=0.0073$). These results do not satisfy either Eq.~(\ref{tool1}) or Eq.~(\ref{tool2}) under the chosen conditions.}
\label{F6sandCR01092}
\end{figure}

To confirm the results obtained from Fig.~\ref{trans} for the switching of dynamics from non-Markovian to Markovian under criterion (\ref{tool2}), we consider the simulated frequency spectrum $|f^{6.0}(\omega)|^{2}$ shown in Fig.~\ref{F6sandCR01092}(a) with $\sigma=0.1092$. The photon dynamics corresponding to this frequency distribution can be determined in the form of Eqs. (\ref{chi1}) and (\ref{chi2}). Figure~\ref{F6sandCR01092}(b) shows the changes of $\alpha_{\chi_{t}}^{6.0}$ and $\alpha_{\chi_{t_2}\chi_{t_1}}^{6.0}$, together with those of $\beta_{\chi_{t}}^{6.0}$ and $\beta_{\chi_{t_2}\chi_{t_1}}^{6.0}$, over time from $t=0$ to $t=160\lambda_{0}$. As shown, $\alpha_{\chi_{t}}^{6.0}$ and $\alpha_{\chi_{t_2}\chi_{t_1}}^{6.0}$ ($\beta_{\chi_{t}}^{6.0}$ and $\beta_{\chi_{t_2}\chi_{t_1}}^{6.0}$) not only monotonically decrease with time, but also have a difference between them of less than 1$\%$ ($\mathscr{N^{\rm{6.0}}_{\alpha}}=0.0099$ and $\mathscr{N^{\rm{6.0}}_{\beta}}=0.0073$). Notably, this result does not satisfy either Eq. (\ref{tool1}) or Eq. (\ref{tool2}). Hence, it can be confirmed that $\chi_{t}^{6.0}$ does indeed switch to Markovian dynamics when $\sigma<0.1093$. It is worth noting that, if the 4 nm (FWHM) filter used in the experimental set-up in Ref. \cite{Liu2011} is replaced with a filter with a FWHM of less than 0.2574 nm, the photon dynamics at $\theta= 6.0^{\circ}$ and $7.5^{\circ}$ are characterized as Markovian by both criterion (\ref{tool1}) and criterion (\ref{tool2}).

\section{Conclusion and Outlook}
This study has investigated the non-Markovianity of the photon dynamics in a birefringent crystal. Process matrices of the polarization state have been faithfully derived from the experimental data reported in the seminal study of Liu \textit{et al.} \cite{Liu2011}. Given this complete description of the dynamical process, one can know the difference between the quantity of quantum-mechanical process in a complete dynamics and that in a process consisting of two sub-processes. It has been shown that the non-Markovian criterion based on this difference can be used to classify non-Markovianity in the dynamical processes that are classified as Markovian by analyzing state properties \cite{BLP, BLP2,RHP,LFS}, or by analyzing the quantity of quantum-mechanical process changing with time. In particular, the photon dynamics which are identified as Markovian in Ref.~\cite{Liu2011} are reliably identified as non-Markovian dynamics by this criterion. The study has also demonstrated the feasibility for switching the photon dynamics from non-Markovian to Markovian using an all-optical set-up. In other words, the results confirm that the condition on the quantity of the quantum-mechanical process can enhance the accuracy and control of quantum system designs for quantum engineering applications.

Future studies may usefully examine whether such a criterion retains the same capability to identify non-Markovianity for general open quantum systems, in which the nonlocal memory effects induced by the initial correlations between local parts of the environment \cite{NonlocalMemoryEffectsintheDynamicsofOQS, nonmarkovianquantumprobe} must also be taken into consideration. In addition, considering the errors induced in experiment (such as the counting statistics and the uncertainty of the tilt angles of the experimental elements \cite{MeasurementofQubits}) in the form of composition and robustness, the accuracy of the criteria can be further improved at the corresponding experimental circumstance.

\acknowledgments
We are grateful to G.-Y. Chen and C.-H. Chou for helpful comments. This work is partially supported by the Ministry of Science and Technology, Taiwan, under Grants No. MOST 104-2112-M-006-016-MY3 and No. MOST 107-2628-M-006-001-MY4.

\appendix*

\section{Other non-Markovian criteria}
\label{othercriteria}

In order to compare the results of identifying the non-Markovianity of the photon dynamics by different non-Markovian criteria, this section commences by briefly introducing the BLP, RHP, and LFS non-Markovian criteria, and then uses these criteria to investigate the constructed process matrices in Eq.~(\ref{chi1}). Using the BLP and RHP criteria, the process matrices derived from the experimental data are proved to be in accordance with the photon dynamics in Ref. \cite{Liu2011}. On the other hand, the difference of the capability to classify non-Markovianity between these criteria and the HCL criteria can be found. (See Sec. \ref{comparison}.)

\begin{figure}[t]
\includegraphics[width=8.6cm]{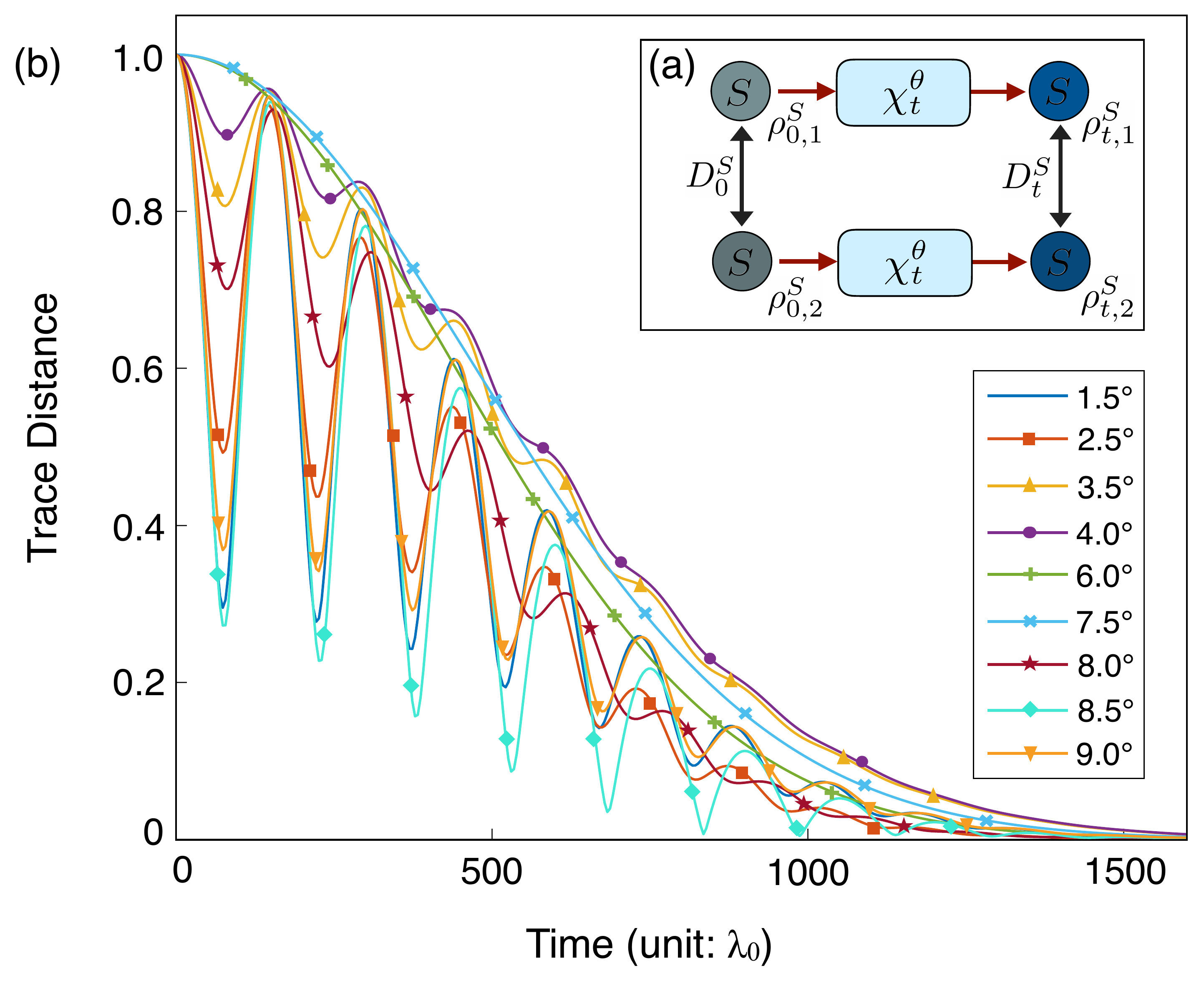}
\caption{Non-Markovianity of the experimental dynamics identified by the BLP non-Markovian criterion. (a) Schematic illustration showing basic concept of the BLP criterion. (b) Identification of the non-Markovianity of $\chi_{t}^{\theta}$ in Eq.~(\ref{chi1}) using the BLP criterion. The quantities of the trace distance when $\theta=6.0^{\circ}$ and $7.5^{\circ}$ monotonically decrease over time, leading to $\mathscr{N}_{\rm{BLP}}=0$. In other words, the photon dynamics $\chi_{t}^{6.0}$ and $\chi_{t}^{7.5}$ are classified as Markovian by the BLP criterion, while the remaining photon dynamics are non-Markovian. Notably, the photon dynamics when $\theta=6.0^{\circ}$ and $7.5^{\circ}$ are experimentally characterized as Markovian in Ref.~\cite{Liu2011}. (See Fig.~\ref{changeofTDandCon} in Sec. \ref{Previous experiment}.)}
\label{resultofBLP}
\end{figure}

\subsection{BLP criterion}
\label{BLPc}
The concept of the BLP criterion is based on the contractive property of the trace distance which denotes the ability to distinguish two different states. Consider, for instance, the trace distance of two initial system states, $\rho_{0,1}^{S}$ and $\rho_{0,2}^{S}$, written as $D_{0}^{S}={ \|\rho_{0,1}^{S}-\rho_{0,2}^{S}\|}/2$, where $\|\rho_{0,1}^{S}-\rho_{0,2}^{S}\|=tr[\sqrt{(\rho_{0,1}^{S}-\rho_{0,2}^{S})^{\dag}(\rho_{0,1}^{S}-\rho_{0,2}^{S})}]$. After undergoing the photon dynamics $\chi_{t}^{\theta}$, the trace distance of the final states can be expressed as $D^{S}_{t}={ \|\chi_{t}^{\theta}(\rho_{0,1}^{S})-\chi_{t}^{\theta}(\rho_{0,2}^{S})\|}/2={ \|\rho_{t,1}^{S}-\rho_{t,2}^{S}\|}/2$.  [See Fig.~\ref{resultofBLP}(a).]

According to the contractive property, any CPTP map $\chi_{t}^{\theta}$ gives rise to a decrease of trace distance \cite{Nielsen&Chuang}. As guaranteed by the semigroup property (see Sec. \ref{MarDynamic}), if $\chi_{t}^{\theta}$ is Markovian, the decomposed subprocesses $\chi_{t_1}^{\theta}$ and $\chi_{t_2}^{\theta}$ are also CPTP maps, leading to $D^{S}_{0}\geq D^{S}_{t_{1}}\geq D^{S}_{t}$ for arbitrary time $t_1$. In other words, if $\chi_{t}^{\theta}$ is divisible (Markovian), the trace distance monotonically decreases over time. Thus, if any increase of the trace distance induced by the sub-processes for arbitrary time interval $t_1$ is observed, the $\chi_{t}^{\theta}$ is said to be non-Markovian. The increased quantity of the trace distance is then defined as the non-Markovianity of $\chi_{t}^{\theta}$ under the BLP criterion, i.e.,
\begin{equation}\label{NBLP}
\mathscr{N}_{\rm{BLP}}=\max\limits_{\rho_{0,1}^{S}, \rho_{0,2}^{S}} \int_{\frac{dD}{dt}>0} dt \frac{d}{dt}D^{S}_{t}.
\end{equation}
The results obtained when identifying the non-Markovianity of the photon dynamics given in Table.~\ref{FreAnalysis} using the BLP criterion are shown in Fig.~\ref{resultofBLP}(b).

It is worth noting that, to implement the BLP criterion, one needs to find an optimal set of the initial states, $\rho_{0,1}^{S}$ and $\rho_{0,2}^{S}$, to reflect the maximum amount of non-Markovianity, whereas the RHP criterion, LFS criterion and HCL criteria are independent of initial states for the dynamical process.

\begin{figure}[t]
\includegraphics[width=8.6cm]{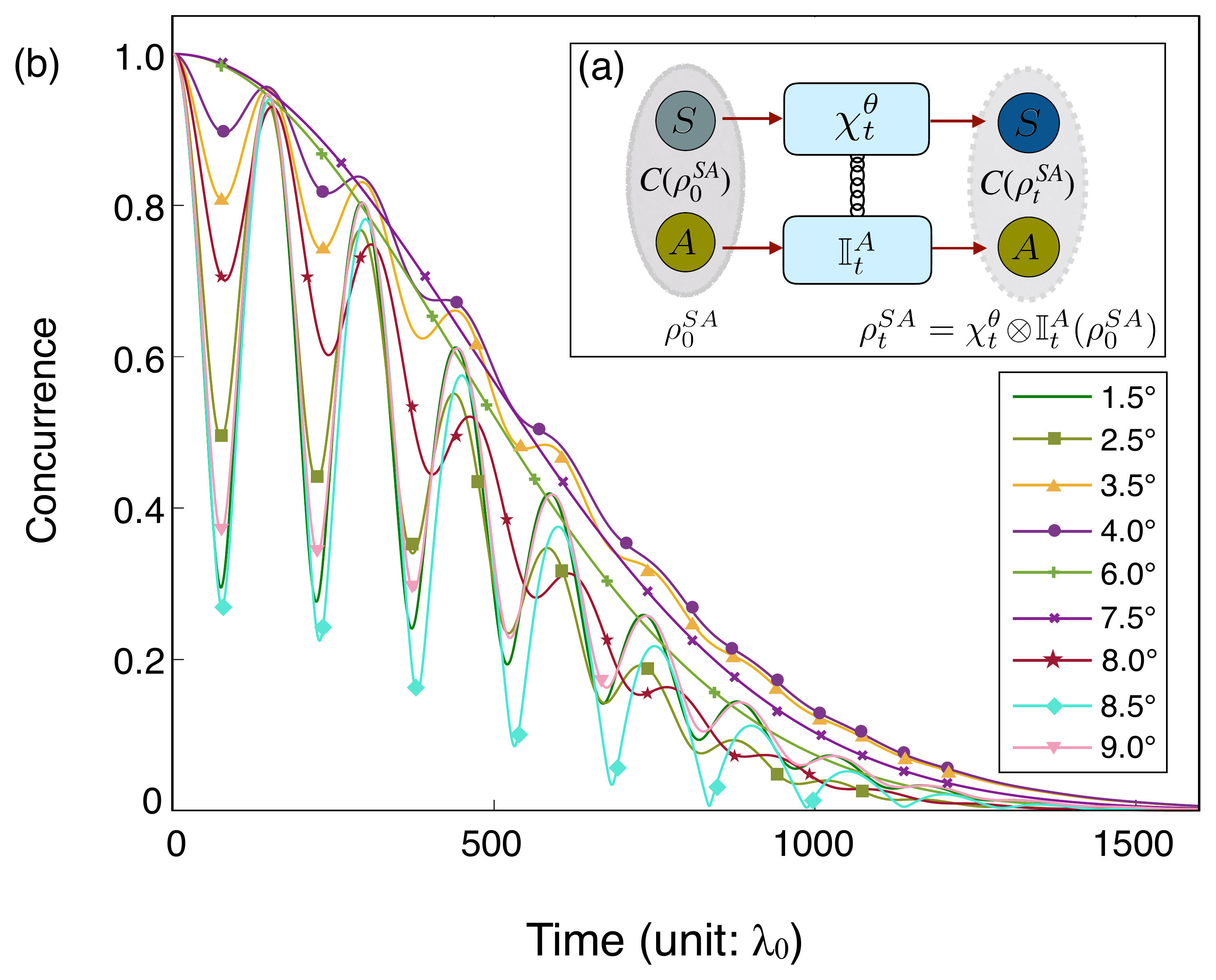}
\caption{Non-Markovianity of the experimental dynamics identified by the RHP non-Markovian criterion. (a) Schematic illustration showing basic concept of the RHP criterion. (b) Identification of the non-Markovianity of $\chi_{t}^{\theta}$ in Eq.~(\ref{chi1}) using the RHP criterion. The amounts of the concurrence when $\theta=6.0^{\circ}$ and $7.5^{\circ}$ monotonically decrease over time. Thus, the photon dynamics $\chi_{t}^{6.0}$ and $\chi_{t}^{7.5}$ are Markovian since $\mathscr{N}_{\rm{RHP}}=0$. The other dynamics are identified as non-Markovian by the RHP criterion. These results are consistent with the experimental findings in \cite{Liu2011}.}
\label{resultofRHP}
\end{figure}

\subsection{RHP criterion}
\label{RHPc}
The RHP criterion is based on the change of intensity of the entanglement between the system and an ancilla, which is kept isolated from the decoherence environment. The intensity of the entanglement can be described in a concurrence form \cite{concurrence}, as illustrated in Fig.~\ref{resultofRHP}(a). The state of the principal system and the ancilla is initially prepared in a maximally entangled state, $\rho^{SA}_{0}=\ket{\psi^{SA}_{0}}\bra{\psi^{SA}_{0}}$, where $\ket{\psi_{0}^{SA}}=(\ket{H^SH^A}+\ket{V^SV^A})/\sqrt{2}$. After the principal system undergoes the photon dynamics $\chi_{t}^{\theta}$, the state of the composite system changes to $\rho^{SA}_{t}=\chi_{t}^{\theta}\otimes\mathbb{I}_{t}^{A}(\rho^{SA}_{0}$), where $\mathbb{I}_{t}^{A}$ denotes an identity process operating on the ancilla. The final state $\rho^{SA}_{t}$ can be measured using two-photon state tomography. The concurrence $C$ of $\rho^{SA}_{0}$ and $\rho^{SA}_{t}$ can then be determined as \cite{concurrence}:
\begin{equation}\label{concurrence}
C(\rho^{SA})=\max(0, \sqrt{e_{1}}-\sqrt{e_{2}}-\sqrt{e_{3}}-\sqrt{e_{4}}),
\end{equation}
where $e_{i}$ are the eigenvalues of $\rho^{SA}(Y\otimes Y){\rho^{SA}}^{*}(Y\otimes Y)$, in which $Y$ is the Pauli matrix and ${\rho^{SA}}^{*}$ is the complex conjugate of $\rho^{SA}$.

Since local CPTP maps $\chi_{t}^{\theta}$ do not increase the correlation between the principal system and the ancilla, any CPTP $\chi_{t_1}^{\theta}$ and $\chi_{t_2}^{\theta}$ decomposed from $\chi_{t}^{\theta}$ with arbitrary time $t_{1}$ do not increase the concurrence when the system undergoes these subprocesses. Furthermore, if $\chi_{t}^{\theta}$ is divisible, i.e., is Markovian, the concurrence monotonically decreases over time, leading to $C(\rho_{0}^{SA})\geq C(\rho_{t_{1}}^{SA})\geq C(\rho_{t}^{SA})$. Hence the increase of the concurrence is then defined as the non-Markovianity of $\chi_{t}^{\theta}$, i.e.,
\begin{equation}\label{NRHP}
  \mathscr{N}_{\rm{RHP}}= \int_{\frac{dC}{dt}>0} dt \frac{d}{dt}C(\rho_{t}^{SA}).
\end{equation}
The results obtained when identifying the non-Markovianity of the photon dynamics given in Table~\ref{FreAnalysis} using the RHP criterion are shown in Fig.~\ref{resultofRHP}(b).

Note that, to use RHP criterion, we first need to prepare the system and the ancilla in a maximally entangled state. If such an initial state is not ideally prepared to be maximally entangled, there exist possible errors in certifying non-Markovianity by the RHP criterion. Compared with the RHP criterion, the HCL criterion (\ref{tool2}) directly use semigroup property to examine whether a process can be decomposed into two subprocesses, which makes it possible to identify non-Markovianity for processes classified as Markovian by the RHP criterion.

\begin{figure}[t]
\includegraphics[width=8.6cm]{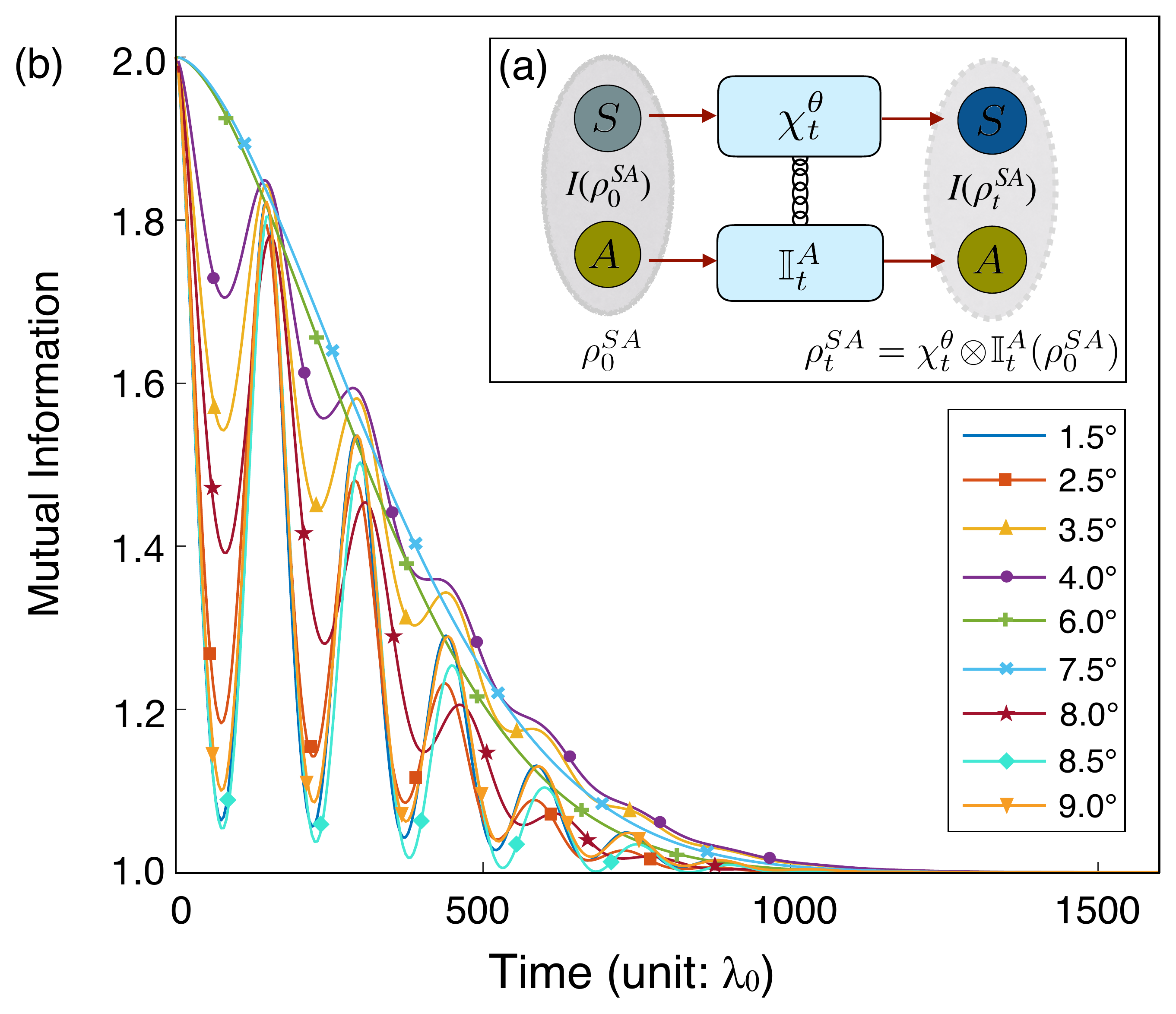}
\caption{Non-Markovianity of the experimental dynamics identified by the LFS non-Markovian criterion. (a) Schematic illustration showing a basic concept of the LFS criterion. (b) Identification of the non-Markovianity of $\chi_{t}^{\theta}$ in Eq.~(\ref{chi1}) using the LFS criterion. The amounts of the mutual information when $\theta=6.0^{\circ}$ and $7.5^{\circ}$ monotonically decrease with time. Thus, the photon dynamics $\chi_{t}^{6.0}$ and $\chi_{t}^{7.5}$ are certified as Markovian. By contrast, the other dynamics are all identified as non-Markovian. Notably, these results are consistent with those obtained using the BLP and RHP criteria.}
\label{resultofLFS}
\end{figure}

\subsection{The LFS criterion}
\label{LFSc}
In employing an ancilla to identify the non-Markovianity of $\chi_{t}^{\theta}$, the underlying concept of the LFS criterion is similar to that of the RHP criterion. However, instead of quantifying the correlations between the principal system and the ancilla by entanglement, the LFS criterion utilizes a mutual information measure to identify non-Markovianity [see Fig.~\ref{resultofLFS}(a)]. The amount of mutual information can be expressed as $I(\rho_{t}^{SA})=S(\rho_{t}^{S})+S(\rho_{t}^{A})-S(\rho_{t}^{SA})$, where $S=-tr[\rho\log_{2}\rho]$ denotes the von Neumann entropy, and $\rho_{t}^{S}=tr_{A}[\rho^{SA}_{t}]$ and $\rho_{t}^{A}=tr_{S}[\rho^{SA}_{t}]$ are the reduced states of the system and ancilla, respectively. It is clear that, if the initial state of the system and the ancilla are not maximally entangled, we cannot faithfully identify the non-Markovianity using the LFS criterion.

From this perspective, the reduction of $I(\rho_{t}^{SA})$ can be interpreted as a loss of information from the principal system into the environment during the Markovian evolution. Thus, if $\chi_{t}^{\theta}$ is Markovian, the reduction of the mutual information is monotonic over time. Conversely, if $I(\rho_{t}^{SA})$ increases, the information flows back from the environment into the principal system. The amount by which the mutual information increases can then be defined as the non-Markovianity of $\chi_{t}^{\theta}$, i.e.,
\begin{equation}\label{NLFS}
  \mathscr{N}_{\rm{LFS}}= \int_{\frac{dI}{dt}>0} dt \frac{d}{dt}I(\rho_{t}^{SA}).
\end{equation}
The results obtained when identifying the non-Markovianity of the photon dynamics given in Table.~\ref{FreAnalysis} using the LFS criterion are shown in Fig.~\ref{resultofLFS}(b).

\end{document}